\documentclass[12pt]{article}%
\usepackage{ amsmath, graphics, rotating}%

\begin{document}
\large
\begin{center}{\large\bf MASSLESS ELEMENTARY PARTICLES
IN A QUANTUM THEORY OVER A GALOIS FIELD}
\end{center}
\vskip 1em \begin{center} {\large F. M. Lev} \end{center}
\vskip 1em \begin{center} {\it (E-mail: felixlev@hotmail.com)} 
\end{center}
\vskip 1em

{\it We consider massless elementary particles in a quantum theory
based on a Galois field (GFQT). We previously showed that 
the theory has a new symmetry 
between particles and antiparticles, which has no analogue 
in the standard approach. We now prove that the symmetry is 
compatible with all
operators describing massless particles. Consequently,
massless elementary particles can have only the
half-integer spin (in conventional units), and
the existence of massless neutral elementary particles
is incompatible with the spin-statistics theorem.
In particular, this implies that the photon and the graviton 
in the GFQT can only be composite particles.}

\begin{flushleft} {\bf Keywords:} Galois fields, massless particles,
modular representations \end{flushleft} 

\section{Introduction}
\label{S1} 
 
At high energies any particle can be 
created and annihilated by other particles in different
reactions. For this reason the property of a particle to be 
elementary or
composite has no clear experimental meaning. However, in theory
this property is well defined. By definition, a particle is
called elementary if the full set of its wave functions forms
a space of irreducible representation (IR) for the symmetry 
group or algebra in the given theory. Such an approach has 
been first proposed by Wigner in Ref. \cite{Wigner} where
unitary IRs of the Poincare group have been constructed.

In the standard approach to quantum theory each elementary 
particle either has or does not have the corresponding
antiparticle with the same mass and spin. The latter case can 
also be treated in such a way that the particle and its 
antiparticle are the same. Elementary particles with such 
a property are called neutral. 

Let us briefly discuss how the standard theory explains the
existence of antiparticles. Consider, for example, the
electron and the positron which are the antiparticles for
each other. The explanation is based on the fact that the
Dirac equation has solutions with both positive and
negative energies. As noted by Dirac (see e.g. his Nobel
lecture \cite{DirNobel}), the existence of the negative
energy solutions represents a difficulty which should be
resolved. In the standard approach the solution is given
in the framework of second quantization such that the
creation and annihilation operators for the positron have
the usual meaning but they enter the quantum Lagrangian 
with the coefficients representing the negative energy
solutions. This is an implementation of the idea that the
creation or annihilation of an antiparticle can be treated 
respectively, as the annihilation or creation of the 
corresponding particle with the negative energy. However,
since negative energies have no direct physical meaning in
the standard theory, this idea is implemented implicitly
rather than explicitly. Note also that the electron and
the positron are described by unitary IRs with positive
energies, but these representations are fully independent.
At the same time, IRs with negative energies are not used
at all. 

In papers \cite{lev1} we have proposed an approach to
quantum theory where the wave functions of the system
under consideration are described by elements of a linear 
space over a Galois field, and the operators of physical 
quantities - by linear operators in this space. A detailed
discussion of this approach has been given in a recent paper
\cite{lev2}. In particular, it has been shown that at some
conditions such a description gives the same predictions as 
the standard approach. It has also been argued that the
description of quantum systems in terms of Galois fields 
is more natural than the standard description in terms of
complex numbers. 

The first obvious conclusion about 
quantum theory based on a Galois field (GFQT) is as
follows: since any Galois field has only a finite number of 
elements, in the GFQT divergencies cannot exist in 
principle, and all operators are automatically well defined.
It is also natural to expect that, since arithmetic of Galois 
field differs from the standard one, the GFQT has some 
properties which have no analog in the standard theory.  

In particular, as shown in Ref. \cite{lev2}, in contrast to 
the standard approach, where a particle and its
antiparticle are described by independent IRs of the symmetry 
group, in the GFQT a particle and its antiparticle are 
described by the same IR of the symmetry algebra. This
automatically explains the existence of antiparticles and
shows that a particle and its antiparticle represent
different states of the same object. As a consequence, the 
GFQT possesses a new symmetry between particles and 
antiparticles, which has no analog in
the standard quantum theory. Also the problem arises
of whether neutral particles can be elementary or only
composite. 

The problem of existence of neutral 
elementary particles is of greatest interest for massless
particles, e.g. for the photon and the graviton. For this
reason, in the present paper (see Sects. \ref{2} and \ref{S3})
we consider the massless case. In Sect. \ref{S4} the 
new symmetry is described in detail, and in Sect. 
\ref{Vacuum} it is shown that the vacuum condition is
consistent only for particles with the half-integer
spin (in conventional units). In Sect. \ref{S5} we prove 
that the symmetry is 
compatible with all the representation operators for
massless particles. In Sect. \ref{S6} it is shown that,
as a consequence, in the massless case the existence
of massless neutral elementary particles in the GFQT 
is incompatible with the standard relation between spin 
and statistics. 

Although the
notion of the Galois field is extremely simple and
elegant, the majority of physicists is not familiar with
this notion. For this reason, in Ref. \cite{lev2} an
attempt has been made to explain the basic facts about
Galois fields by using arguments which, hopefully, can 
be accepted by physicists. 
The readers who are not familiar with Galois fields can 
also obtain basic knowledge from the standard textbooks
(see e.g. Refs. \cite{VDW}). 
 
\section{Representation operators of the anti de Sitter algebra}
\label{S2}

If a conventional quantum theory has a symmetry group (or
algebra), then there exists a unitary representation of 
the group (or a representation of the algebra by Hermitian
operators) in the Hilbert space describing the quantum 
system under consideration. In the present paper we assume
that the symmetry algebra is the Galois field analog of the 
anti de Sitter (AdS) algebra so(2,3), and quantum systems are
described by representations of this algebra in spaces over
a Galois field (see Ref. \cite{lev2} for details). The 
standard AdS group
is ten-parametric, as well as the Poincare group. However, 
in contrast to the Poincare group, all the representation 
generators are angular momenta. In Ref. \cite{lev2} we 
explained the reason why for our
purposes it is convenient to work with the units 
$\hbar/2=c=1$. Then the representation generators are
dimensionless, and the commutation relations for them can be 
written in the form
\begin{equation}
[M^{ab},M^{cd}]=-2i (g^{ac}M^{bd}+g^{bd}M^{cd}-
g^{ad}M^{bc}-g^{bc}M^{ad})
\label{1}
\end{equation}
where $a,b,c,d$ take the values 0,1,2,3,5, and the operators 
$M^{ab}$ are antisymmetric. The diagonal metric tensor has the 
components $g^{00}=g^{55}=-g^{11}=-g^{22}=-g^{33}=1$.  
In these units the spin of fermions is odd, and the spin of 
bosons is even. If $s$ is the particle spin then the 
corresponding IR of the su(2) algebra has the dimension
$s+1$. Note that if $s$ is interpreted in such a way then it
does not depend on the choice of units (in contrast to the
maximum eigenvalue of the $z$ projection of the spin
operator).    

For analyzing IRs implementing Eq. (1), it is convenient to
work with another set of ten operators. Let $(a_j',a_j",h_j)$ 
$(j=1,2)$ be two independent sets of operators satisfying the 
commutation relations 
\begin{equation}
[h_j,a_j']=-2a_j'\quad [h_j,a_j"]=2a_j"\quad [a_j',a_j"]=h_j
\label{2}
\end{equation}
The sets are independent in the sense that
for different $j$ they mutually commute with each other. 
We denote additional four operators as $b', b",L_+,L_-$.
The meaning of $L_+,L_-$ is as follows. The operators 
$L_3=h_1-h_2,L_+,L_-$ satisfy the commutation relations
of the su(2) algebra
\begin{equation}
[L_3,L_+]=2L_+\quad [L_3,L_-]=-2L_-\quad [L_+,L_-]=L_3
\label{3}
\end{equation}
while the other commutation relations are as follows
\begin{eqnarray}
&[a_1',b']=[a_2',b']=[a_1",b"]=[a_2",b"]=\nonumber\\
&[a_1',L_-]=[a_1",L_+]=[a_2',L_+]=[a_2",L_-]=0\nonumber\\
&[h_j,b']=-b'\quad [h_j,b"]=b"\quad 
[h_1,L_{\pm}]=\pm L_{\pm},\nonumber\\
&[h_2,L_{\pm}]=\mp L_{\pm}\quad [b',b"]=h_1+h_2\nonumber\\
&[b',L_-]=2a_1'\quad [b',L_+]=2a_2'\quad [b",L_-]=-2a_2"\nonumber\\
&[b",L_+]=-2a_1",\quad [a_1',b"]=[b',a_2"]=L_-\nonumber\\
&[a_2',b"]=[b',a_1"]=L_+,\quad [a_1',L_+]=[a_2',L_-]=b'\nonumber\\
&[a_2",L_+]=[a_1",L_-]=-b"
\label{4}
\end{eqnarray}  
At first glance, these relations might seem 
rather chaotic, but they are in fact
very natural in the Weyl basis of the so(1,4) algebra.

The relation between the above sets of ten operators is as follows
\begin{eqnarray}
&M_{10}=i(a_1"-a_1'-a_2"+a_2')\quad M_{15}=a_2"+a_2'-a_1"-a_1'\nonumber\\
&M_{20}=a_1"+a_2"+a_1'+a_2'\quad M_{25}=i(a_1"+a_2"-a_1'-a_2')\nonumber\\
&M_{12}=L_3\quad M_{23}=L_++L_-\quad M_{31}=-i(L_+-L_-)\nonumber\\
&M_{05}=h_1+h_2\quad M_{35}=b'+b"\quad M_{30}=-i(b"-b')
\label{5}
\end{eqnarray}
In addition, if $^*$ is used to denote the Hermitian conjugation,
$L_+^*=L_-$, $a_j^{'*}=a_j"$, $b^{'*}=b"$ and $h_j^*=h_j$ then
the operators $M^{ab}$ are Hermitian (we do not discuss the
difference between selfadjoined and Hermitian operators).   

Let $p$ be a prime number, and $F_{p^2}$ be a Galois field
containing $p^2$ elements. This field has only one
nontrivial automorphism $a\rightarrow {\bar a}$ (see e.g.
Refs. \cite{VDW,lev2}) which is the analog of complex
conjugation in the field of complex numbers. The automorphism
can be defined as $a\rightarrow {\bar a}= a^p$ \cite{VDW}. 
Our goal is to find IRs, implementing the commutation relations 
(\ref{2}-\ref{4}) in
spaces over $F_{p^2}$. Representations in spaces over fields
of nonzero characteristics are called modular representations.
A review of the theory of modular IRs can be found e.g. in
Ref. \cite{FrPa}. In the present paper we do not need a
general theory since modular IRs in question can be
constructed explicitly. A modular analog of the Hilbert
space is a linear space $V$ over $F_{p^2}$ supplied by a
scalar product (...,...) such that for any $x,y\in V$
and $a\in F_{p^2}$, $(x,y)\in F_{p^2}$
and the following properties are satisfied: 
\begin{equation}
(x,y) =\overline{(y,x)},\quad (ax,y)=\bar{a}(x,y),\quad 
(x,ay)=a(x,y)
\label{6}
\end{equation}   
In the modular case $^*$ is used to denote the Hermitian
conjugation, such that $(Ax,y)=(x,A^*y)$.

Eq. (\ref{2}) defines the commutation relations for 
representations of the sp(2) algebra. These representations
play an important role in constructing modular IRs of the
so(2,3) algebra. For this reason, following Refs. 
\cite{lev1,lev2}, we describe below modular IRs of the
sp(2) algebra such that the representation generators are
denoted as $a',a",h$. 

The  Casimir operator of the second order for the algebra
(\ref{2}) has the form
\begin{equation}
K=h^2-2h-4a"a'=h^2+2h-4a'a"
\label{7}
\end{equation}
We will consider representations with the vector $e_0$, such that
\begin{equation}
a'e_0=0,\quad he_0=q_0e_0,\quad (e_0,e_0)=1 
\label{8}
\end{equation} 
One can easily prove \cite{lev1,lev2} that $q_0$ is
"real", i.e. $q_0\in F_p$ where $F_p$ is the residue field
modulo $p$: $F_p=Z/Zp$ where $Z$ is the ring of integers.
The field $F_p$ consists of $p$ elements and represents the
simplest possible Galois field.

Denote $e_n =(a")^ne_0$.
Then it follows from Eqs. (\ref{7}) and (\ref{8}),
that for any $n=0,1,2,...$
\begin{equation}
he_n=(q_0+2n)e_n,\quad Ke_n=q_0(q_0-2)e_n, 
\label{9}
\end{equation} 
\begin{equation}
a'a"e_n=(n+1)(q_0+n)e_n
\label{10}
\end{equation}
\begin{equation}
(e_{n+1},e_{n+1})=(n+1)(q_0+n)(e_n,e_n)
\label{11}
\end{equation}

The case $q_0=0$ is trivial and corresponds to zero representation,
so we assume that $q_0\neq 0$. Then we have the case when ordinary 
and modular representations considerably differ each other. 
Consider first the ordinary case when $q_0$ is any real positive 
number. Then IR is infinite-dimensional, $e_0$ is a vector
with a minimum eigenvalue of the operator $h$ (minimum weight)
and there are no vectors with the maximum weight. This is in 
agreement with the well known fact that unitary IRs of
noncompact groups are infinite dimensional. However in the
modular case $q_0$ is one of the numbers $1,...p-1$. 
The set $(e_0,e_1,...e_N)$ will be a basis
of IR if $a"e_i\neq 0$ for $i<N$ and $a"e_N=0$. These conditions
must be compatible with $a'a"e_N=0$. Therefore, as follows from
Eq. (\ref{11}), $N$ is defined by the condition $q_0+N=0$ in
$F_p$. As a result, $N=p-q_0$ and the dimension of IR is equal
to $p-q_0+1$.  

One might say that $e_0$ is the vector with the
minimum weight while $e_N$ is the vector with the maximum weight.
However the notions of "less than" or "greater than" have only
a limited sense in $F_p$, as well as the notion of positive
and negative numbers in $F_p$. If $q_0$ is positive in this sense
(see Ref. \cite{lev2} for details), then Eqs. (\ref{8}) and
(\ref{9}) indicate that the modular IR under consideration can be 
treated as the modular analog of IR with "positive energies".
However it is easy to see that $e_N$ is the
eigenvector of the operator $h$ with the eigenvalue $-q_0$ in
$F_p$, and the same IRs can be treated as the modular analog
of IRs with "negative energies" (see Ref. \cite{lev2} for
details). 

\section{Massless modular representations of the AdS algebra}
\label{S3}

There exists a vast literature on ordinary IRs of the so(2,3) 
algebra in Hilbert spaces. The representations relevant for 
elementary particles in the AdS space have been constructed 
for the first time in Refs. \cite{Fronsdal,Evans}, while 
modular representations of algebra (\ref{2}-\ref{4}) have 
been investigated for the first time by Braden \cite{Braden}. 
In Refs. \cite{lev1,lev2} we have reformulated his 
investigation in such a way that the correspondence between 
modular and ordinary IRs are straightforward. Our construction
is described below. 

We use the basis in which the operators 
$(h_j,K_j)$ $(j=1,2)$ are diagonal. Here $K_j$ is the 
Casimir operator (\ref{7}) for algebra $(a_j',a_j",h_j)$. 
By analogy with Refs. \cite{Evans,Braden} we introduce the
operators 
\begin{eqnarray}
&B^{++}=b"-a_1"L_-(h_1-1)^{-1}-a_2"L_+(h_2-1)^{-1}+\nonumber\\
&a_1"a_2"b'[(h_1-1)(h_2-1)]^{-1}\quad 
B^{+-}=L_+-a_1"b'(h_1-1)^{-1}\nonumber\\
&B^{-+}=L_--a_2"b'(h_2-1)^{-1}\quad B^{--}=b'
\label{12}
\end{eqnarray}
and consider their action only on the space of "minimal" 
sp(2)$\times$sp(2) vectors, i.e. such vectors $x$ that 
$a_j'x=0$ for $j=1,2$, and $x$ is the eigenvector of the
operators $h_j$. 

It is easy to see that if $x$ is a minimal vector such that
$h_jx=\alpha_jx$ then $B^{++}x$ is the minimal eigenvector of the
operators $h_j$ with the eigenvalues $\alpha_j+1$, $B^{+-}x$ - 
with the eigenvalues $(\alpha_1+1,\alpha_2-1)$, 
$B^{-+}x$ - with the eigenvalues $(\alpha_1-1,\alpha_2+1)$, 
and $B^{--}x$ - with the eigenvalues $\alpha_j-1$.

By analogy with the construction of ordinary representations with
positive energy \cite{Fronsdal,Evans}, we require the existence
of the vector $e_0$ satisfying the conditions
\begin{eqnarray}
&a_j'e_0=b'e_0=L_+e_0=0\quad h_je_0=q_je_0\nonumber\\
&(e_0,e_0)=1\quad (j=1,2)
\label{13}
\end{eqnarray}
In the ordinary case the massless IRs are characterized by
the condition $q_2=1$. In the modular case we have the
same condition but now $q_2\in F_p$.

It is well known that $M^{05}=h_1+h_2$ is the AdS analog of 
the energy operator, since $M^{05}/2R$ becomes the usual 
energy when the AdS group is contracted to the Poincare one 
(here $R$ is the radius of the AdS space). As follows from 
Eqs. (\ref{2}) and (\ref{4}), the operators $(a_1',a_2',b')$ 
reduce the AdS energy by two units. Therefore in the 
conventional theory
$e_0$ is the state with the minimum energy. In this theory 
the spin in our units is equal to the maximum value of the 
operator $L_3=h_1-h_2$ in the
"rest state". For this reason we use $s$ to denote $q_1-q_2$.
In our units $s=2$ for the photon, and $s=4$ for the graviton.
Note that in contrast to the Poincare invariant theories, 
massless particles in the AdS case do have states which can be
treated as rest ones (see below).

The problem arises how to define the action of the operators 
$B^{++}$ and $B^{-+}$ on $e_0$ which is the 
eigenvector of the operator $h_2$ with the eigenvalue $q_2=1$.
A possible way to resolve ambiguities 0/0 in matrix elements 
is to write $q_2$ in the form $q_2=1+\epsilon$ and take the 
limit $\epsilon\rightarrow 0$ at the
final stage of computations. This confirms a well known fact that
analytical methods can be very useful in problems involving only
integers. At the same time, one can justify the results
by using only integers (or rather elements of the Galois field
in question), but we will not go into details. 

By using the above prescription, we require that
\begin{eqnarray}
B^{++}e_0=[b"-a_1"L_-(h_1-1)^{-1}]e_0\quad B^{-+}e_0=L_-e_0
\label{14}
\end{eqnarray}
if $s\neq 0$ (and thus $h_1\neq 1$), and
\begin{equation}
B^{++}e_0=b"e_0\quad B^{+-}e_0=B^{-+}e_0=0
\label{15}
\end{equation}
if $s=0$. As follows from the previous remarks, so defined operators
transform minimal vectors to minimal ones, and therefore the element
\begin{equation}
e_{nk}=(B^{++})^n(B^{-+})^ke_0
\label{16}
\end{equation}
is the minimal sp(2)$\times$sp(2) vector with the 
eigenvalues of the operators $h_1$ and $h_2$ equal to 
$q_1+n-k$ and $q_2+n+k$, respectively.

One can directly verify that, as follows from Eqs. 
(\ref{2}-\ref{4})
\begin{eqnarray}
&B^{-+}B^{++}(h_1-1)=B^{++}B^{-+}(h_1-2)\nonumber\\
&B^{+-}B^{++}(h_2-1)=B^{++}B^{+-}(h_2-2),
\label{17}
\end{eqnarray}
and, in addition, as follows from Eq. (\ref{13}) (see
Ref. \cite{lev2} for details)
\begin{equation}
B^{--}e_{nk}=a(n,k)e_{n-1,k}\quad 
B^{+-}e_{nk}=b(n,k)e_{n,k-1}
\label{18}
\end{equation}
where
\begin{eqnarray}
&a(n,k)=\frac{n(q_1+q_2+n-3)(q_1+n-1)(q_2+n-2)}
{(q_1+n-k-2)(q_2+n+k-2)}\nonumber\\
&b(n,k)=\frac{k(s+1-k)(q_2+k-2)}{q_2+n+k-2}
\label{19}
\end{eqnarray}
As follows from these expressions, the elements $e_{nk}$ form
a basis in the space of minimal sp(2)$\times$sp(2) vectors,
and our next goal is to determine the range of the numbers $n$
and $k$.

Consider first the quantity $b(0,k)=k(s+1-k)$ and let 
$k_{max}$ be the maximum value of $k$. For consistency
we should require that if $k_{max}\neq 0$ then $k=k_{max}$ is
the greatest value of $k$ such that $b(0,k) \neq 0$ for
$k=1,...k_{max}$. We conclude that $k$ can take only the 
values of $0,1,..s$. 

Let now $n_{max}(k)$ be the maximum value of $n$ at a given
$k$. For consistency we should require that if 
$n_{max}(k) \neq 0$ then $n_{max}(k)$ is the greatest value
of $n$ such that $a(n,k)\neq 0$ for $n=1,...n_{max}(k)$.
As follows from Eq. (\ref{19}), in the massless case
(when $q_2=1$) $a(1,k)=0$ for $k=1,..s-1$ if such values of
$k$ exist (i.e. when $s\geq 2$), and $a(n,k)=n(s+n)$ if
$k=0$ or $k=s$. We conclude that at $k=1,...s-1$, the
quantity $n$ can take only the value $n=0$ while at $k=0$
or $k=s$, the possible values of $n$ are $0,1,...n_{max}$
where $n_{max}=p-s-1$ (in contrast to the standard theory
where $n=0,1,...\infty$).

The full basis of the representation space can be chosen in the
form 
\begin{equation}
e(n_1n_2nk)=(a_1")^{n_1}(a_2")^{n_2}e_{nk} 
\label{20}
\end{equation}
where, as follows from the
results of the preceding section, 
\begin{eqnarray}
&n_1=0,1,...N_1(n,k)\quad n_2=0,1,...N_2(n,k)\nonumber\\
&N_1(n,k)=p-q_1-n+k\quad N_2(n,k)=p-q_2-n-k
\label{21}
\end{eqnarray}

We conclude that, in contrast with the standard theory,
where IRs of Lie algebras by Hermitian operators are
necessarily infinite-dimensional, massless modular IRs
are finite-dimensional and even finite since the field
$F_{p^2}$ is finite. This is in agreement with a
general statement proved by Zassenhaus
\cite{Zass} that any modular IR is finite-dimensional. 

Let us now discuss why IRs in question can be
treated as massless. It is easy to see that
\begin{eqnarray}
&h_1e(n_1n_2nk)=(q_1+n-k+2n_1)e(n_1n_2nk)\nonumber\\
&h_2e(n_1n_2nk)=(q_2+n+k+2n_2)e(n_1n_2nk)\nonumber\\
&M^{05}e(n_1n_2nk)=(q_1+q_2+2n+2n_1+2n_2)e(n_1n_2nk)
\label{22}
\end{eqnarray}
Therefore in the standard AdS theory the
corresponding IR is characterized by the minimum
AdS energy equal to $q_1+q_2=2q_2+s$. Since in the
usual case the mass is treated as the minimum energy,
and the conventional energy is equal to $M^{05}/2R$,
the conventional mass becomes zero when $q_2=1$ and
$R\rightarrow \infty$. However this observation is 
still insufficient to conclude that $q_2=1$ is
distinguished among other values of $q_2$ since 
$(2q_2+s)/2R\rightarrow 0$ when $R\rightarrow \infty$
if $q_2$ is any finite number.
Let us recall that massless particles in conventional 
theory do not have "rest states", and for this reason
the value of $s$ does not characterize the number of
states in the corresponding IR of the su(2) algebra.
Instead, massless particles are characterized by
helicity which can have only two values: $s$ or $-s$. 
The AdS analog of this situation is that at $q_2=1$
and $n>0$ there exist only the elements $e_{nk}$ with
$k=0$ and $k=s$. Only if $n=0$, there exist the elements
$e_{nk}$ with $k=0,1...s$. When the AdS algebra is
contracted to the Poincare one (the meaning of contruction
is well known \cite{IW}), the discrete spectrum becomes
the continuous one, and the probability for a particle
to have zero energy is negligible.

Taking into consideration the above remarks, in the
literature the massless case is often characterized not
only by the condition $q_2=1$, but also by the condition
$s\geq 2$, since only in that case $1\leq s-1$. We
assume only that $q_2=1$ while the spin can be 
arbitrary.

The above results can be summarized in the expressions
\begin{eqnarray}
&B^{++}e_{nk}=e_{n+1,k}\quad (k=0,s;\,\, n=0,1...n_{max}-1)\nonumber\\
&B^{--}e_{nk}=n(s+n)e_{n-1,k}\quad (k=0,s;\,\, n=1,...n_{max})\nonumber\\
&B^{+-}e_{0k}=k(s+1-k)e_{0,k-1}\quad (k=1,...s)\nonumber\\
&B^{-+}e_{0k}=e_{0,k+1}\quad (k=0,1,..s-1)
\label{23}
\end{eqnarray}
while at other values of $n$ and $k$ the action of these
operators on $e_{nk}$ is equal to zero.
 
Our next task is to compute the quantities 
$$Norm(n_1n_2nk)=(e(n_1n_2nk),e(n_1n_2nk)).$$
By using Eqs. (\ref{11}) and (\ref{23}) one can show that
\begin{eqnarray}
&Norm(n_1n_2nk)=F(n_1n_2nk)G(k), \quad where \nonumber\\
&F(n_1n_2nk)=n_1!n_2!(n_1+n+s-k)!(n+n_2+k)!\nonumber\\
&G(k)=s!/[(s-k)!]^2
\label{24}
\end{eqnarray}

In standard Poincare and AdS theories there also exist IRs with
negative energies (as noted in Sect. \ref{S1}, they are not
used in the standard approach).
They can be constructed by analogy with positive energy IRs.
Instead of Eq. (\ref{13}) one can require the existence of the
vector $e_0'$ such that
\begin{eqnarray}
&a_j"e_0'=b"e_0'=L_-e_0'=0\quad h_je_0'=-q_je_0'\nonumber\\
&(e_0',e_0')\neq 0\quad (j=1,2)
\label{25}
\end{eqnarray}
where the quantities $q_1,q_2$ are the same as for positive
energy IRs. It is obvious that positive and negative energy
IRs are fully independent since the spectrum of the operator
$M^{05}$ for such IRs is positive and negative, respectively.
At the same time, as shown in Ref. \cite{lev2},
{\it the modular analog of a positive energy IR 
characterized by $q_1,q_2$ in Eq. (\ref{13}), and the modular 
analog of a negative energy IR characterized by the same 
values of $q_1,q_2$ in Eq. (\ref{25}) represent the same
modular IR.} Since this is the crucial difference between the
standard quantum theory and the GFQT, we give below the
proof (which slightly differs from that in Ref. \cite{lev2}).

Let $e_0$ be a vector satisfying Eq. (\ref{13}). Denote 
$N_1=p-q_1$ and $N_2=p-q_2$. We will prove that the vector 
$x=(a_1")^{N_1}(a_2")^{N_2}e_0$ satisfies the conditions
(\ref{25}), i.e. $x$ can be identified with $e_0'$. 

As follows from Eq. (\ref{9}), the definition of $N_1,N_2$
and the results of the preceding section, the vector $x$
is the eigenvector of the operators $h_1$ and $h_2$ with
the eigenvalues $-q_1$ and $-q_2$, respectively, and, in
addition, it satisfies the conditions $a_1"x=a_2"x=0$.

Let us now prove that $b"x=0$. Since $b"$ commutes with the
$a_j"$, we can write $b"x$ in the form
\begin{equation}
b"x = (a_1")^{N_1}(a_2")^{N_2}b"e_0
\label{26}
\end{equation}
As follows from Eqs. (\ref{4}) and (\ref{13}), 
$a_2'b"e_0=L_+e_0=0$ and $b"e_0$ is the eigenvector
of the operator $h_2$ with the eigenvalue $q_2+1$.
Therefore, $b"e_0$ is the minimal vector of the sp(2)
representation which has the dimension $p-q_2=N_2$.
Therefore $(a_2")^{N_2}b"e_0=0$ and $b"x=0$.

The next stage of the proof is to show that $L_-x=0$.
As follows from Eq. (\ref{4}) and the definition of
$x$,
\begin{equation}
L_-x = (a_1")^{N_1}(a_2")^{N_2}L_-e_0-
N_1(a_1")^{N_1-1}(a_2")^{N_2}b"e_0
\label{27}
\end{equation}
We have already shown that $(a_2")^{N_2}b"e_0=0$,
and therefore it suffice to prove that the first term
in the r.h.s. of Eq. (\ref{27}) is equal to zero. As follows
from Eqs. (\ref{4}) and (\ref{13}), $a_2'L_-e_0=b'e_0=0$,
and $L_-e_0$ is the eigenvector of the operator $h_2$ with the
eigenvalue $q_2+1$. Therefore $(a_2")^{N_2}L_-e_0=0$ and we
have proved that $L_-x=0$.

The fact that $(x,x)\neq 0$ immediately follows from the
definition of the vector $x$ and the results of the preceding
section. Therefore the vector $x$ can be indeed identified with
$e_0'$ and the above statement is proved.

The matrix elements of the operator $M^{ab}$ are defined as
\begin{equation}
M^{ab}e(n_1n_2nk)=\sum_{n_1'n_2'n'k'}
M^{ab}(n_1'n_2'n'k',n_1n_2nk)e(n_1'n_2'n'k')
\label{28}
\end{equation}
In the modular case the trace of each operator $M^{ab}$ is
equal to zero. For the operators $(a_j',a_j",L_{\pm},b',b")$
this is clear immediately: since they do not contain
nonzero diagonal elements at all, they necessarily change
one of the quantum numbers $(n_1n_2nk)$. The proof for the
diagonal operators $h_1$ and $h_2$ is as follows. For each 
IR of the sp(2) algebra with the minimal weight $q_0$ and the 
dimension $N+1$, the eigenvalues of the operator $h$ are 
$(q_0,q_0+2,...q_0+2N)$. The sum of these eigenvalues is
equal to zero in $F_p$ since $q_0+N=0$ in $F_p$ (see the
preceding section). Therefore we conclude that
\begin{equation}
\sum_{n_1n_2nk} M^{ab}(n_1n_2nk,n_1n_2nk)=0
\label{29}
\end{equation} 
This property is very important for investigating a new
symmetry between particles and antiparticles in the GFQT
(see Sect. \ref{S5}).

\section{New symmetry between particles and antiparticles
in GFQT}
\label{S4}

Since $(n_1n_2nk)$ is the complete set of quantum numbers
for the elementary particle in question, we can define 
operators describing annihilation and creation of the 
particle in the
states with such quantum numbers. Let $a(n_1n_2nk)$ be the
operator of particle annihilation in the state described
by the vector $e(n_1n_2nk)$. Then the adjoint operator
$a(n_1n_2nk)^*$ has the meaning of particle creation in
that state. Since we do not normalize the states 
$e(n_1n_2nk)$ to one (see the discussion in Ref. 
\cite{lev2}), we require that the operators $a(n_1n_2nk)$ 
and $a(n_1n_2nk)^*$ should satisfy
either the anticommutation relations
\begin{eqnarray}
&\{a(n_1n_2nk),a(n_1'n_2'n'k')^*\}=\nonumber\\
&Norm(n_1n_2nk)
\delta_{n_1n_1'}\delta_{n_2n_2'}\delta_{nn'}\delta_{kk'}
\label{30}
\end{eqnarray}
or the commutation relation
\begin{eqnarray}
&[a(n_1n_2nk),a(n_1'n_2'n'k')^*]=\nonumber\\
&Norm(n_1n_2nk)
\delta_{n_1n_1'}\delta_{n_2n_2'}\delta_{nn'}\delta_{kk'}
\label{31}
\end{eqnarray}
Then, taking into account the fact that the matrix elements
satisfy the proper commutation relations, it is easy to 
demonstrate that
the operators $M^{ab}$ in the secondly quantized form
\begin{eqnarray}  
&M^{ab}=\sum \{ M^{ab}(n_1'n_2'n'k',n_1n_2nk)\nonumber\\
&a(n_1'n_2'n'k')^*a(n_1n_2nk)/Norm(n_1n_2nk) \}
\label{32}
\end{eqnarray}
satisfy the commutation relations in the form (\ref{1}) or
(\ref{2}-\ref{4}) if the $(a,a^*)$ operators satisfy
either Eq. (\ref{30}) or Eq. (\ref{31}). Here and
henceforth we use a convention that summation over
repeated indices is implied.

In the standard theory, where the particle and its 
antiparticle are described by independent IRs, Eq. 
(\ref{32}) describes either the quantized field for
particles or antiparticles. To be precise, let us assume 
that the operators $a(n_1n_2nk)$ and $a(n_1n_2nk)^*$ are related
to particles while the operators $b(n_1n_2nk)$ and
$b(n_1n_2nk)^*$ satisfy the analogous commutation relations
and describe the annihilation and creation of antiparticles.
Then in the standard theory the operators of the quantized
particle-antiparticle field are given by
\begin{eqnarray}  
&M_{standard}^{ab}=\sum \{M_{particle}^{ab}
(n_1'n_2'n'k',n_1n_2nk)\nonumber\\
&a(n_1'n_2'n'k')^*a(n_1n_2nk)/Norm(n_1n_2nk)\}+\nonumber\\
&\sum \{M_{antiparticle}^{ab}(n_1'n_2'n'k',n_1n_2nk)\nonumber\\
&b(n_1'n_2'n'k')^*b(n_1n_2nk)/Norm(n_1n_2nk)\}
\label{33}
\end{eqnarray}
where the quantum numbers $(n_1n_2nk)$ in each sum take the
values allowable for the corresponding IR. 

In contrast to the standard theory, Eq. (\ref{32}) describes
the quantized field for particles and antiparticles 
simultaneously. When the values of $(n_1n_2n)$ are much less
than $p$, the contribution of such values correctly describes
particles (see Ref. \cite{lev2}) for details). The problem 
arises whether this expression correctly describes 
the contribution of antiparticles in the GFQT. Indeed, 
when the AdS energy is negative, the operator $a(n_1n_2nk)$ 
cannot be treated as the annihilation operator and 
$a(n_1n_2nk)^*$ cannot be treated as the creation
operator. 

Let us recall (see Sect. \ref{S3}) that at any fixed
values of $n$ and $k$, the quantities $n_1$ and $n_2$
can take only the values $0,1...N_1(n,k)$ and
$0,1...N_2(n,k)$, respectively (see Eq. (\ref{21})).
We use $Q_1(n,k)$ and $Q_2(n,k)$ to denote $q_1+n-k$
and $q_2+n+k$, respectively. Then, as follows from Eq.
(\ref{22}), the element $e(n_1n_2nk)$ is the eigenvector
of the operators $h_1$ and $h_2$ with the eigenvalues
$Q_1(n,k)+2n_1$ and $Q_2(n,k)+2n_2$, respectively. 
As follows from the results of Sect. \ref{S2}, the first IR
of the sp(2) algebra has the dimension $N_1(n,k)+1$ and the 
second IR has the dimension $N_2(n,k)+1$. If $n_1=N_1(n,k)$ 
then it follows from Eq. (\ref{22}) that the first 
eigenvalue is equal to $-Q_1(n,k)$ in $F_p$, and if 
$n_2=N_2(n,k)$ then 
the second eigenvalue is equal to $-Q_2(n,k)$ in $F_p$.
We use ${\tilde n}_1$ to denote $N_1(n,k)-n_1$ and 
${\tilde n}_2$ to denote 
$N_2(n,k)-n_2$. Then it follows from Eq. (\ref{22}) that
$e({\tilde n}_1{\tilde n}_2nk)$ is the 
eigenvector  
of the operator $h_1$ with the eigenvalue $-(Q_1(n,k)+2n_1)$ 
and the 
eigenvector of the operator $h_2$ with the eigenvalue 
$-(Q_2(n,k)+2n_2)$. 

In the GFQT the operators $b(n_1n_2nk)$ and 
$b(n_1n_2nk)^*$ cannot be independent of
$a(n_1n_2nk)$ and $a(n_1n_2nk)^*$. The meaning of
the operators $b(n_1n_2nk)$ and $b(n_1n_2nk)^*$
should be such that if the values of $(n_1n_2n)$ are
much less than $p$, these operators can be interpreted
as those describing the annihilation and creation of
antiparticles. Therefore it is reasonable to think that the 
operator $b(n_1n_2nk)$ should be defined in such a way that
it is proportional to $a({\tilde n}_1,{\tilde n}_2,n,k)^*$
and $b(n_1n_2nk)^*$ should be defined in such a way that
it is proportional to $a({\tilde n}_1,{\tilde n}_2,n,k)$. 
In this way we can directly implement the idea that the
creation of the antiparticle with the positive energy
can be described as the annihilation of the particle with the
negative energy, and the annihilation of the antiparticle 
with the positive energy can be described as the creation of 
the particle with the negative energy. As noted in Sect.
\ref{S1}, in the standard theory this idea is implemented 
implicitly.   

As follows from the well known Wilson
theorem $(p-1)!=-1$ in $F_p$ (see e.g. \cite{VDW})
and Eq. (\ref{24})
\begin{equation}
F(n_1n_2nk)F({\tilde n}_1{\tilde n}_2nk) = (-1)^s
\label{34}
\end{equation}   
We now {\bf define} the $b$-operators as follows. 
\begin{equation}
a(n_1n_2nk)^*=\eta(n_1n_2nk) b({\tilde n}_1{\tilde n}_2nk)/
F({\tilde n}_1{\tilde n}_2nk) 
\label{35}
\end{equation}
where $\eta(n_1n_2nk)$ is some function. 
Note that in the standard theory the 
CPT-transformation in Schwinger's formulation transforms 
$a^*$ to $b$ (see e.g. Refs. \cite{AB,Wein}), but in 
that case the 
both operators refer only to positive energies, in contrast to 
Eq. (\ref{35}). In contrast to the standard CPT-transformation,
where the sets $(a,a^*)$ and $(b,b^*)$ are independent, Eq.
(\ref{35}) represents not a transformation but a definition.

As a consequence of this definition, 
\begin{eqnarray}
&a(n_1n_2nk)=\bar{\eta}(n_1n_2nk) b({\tilde n}_1{\tilde n}_2nk)^*/
F({\tilde n}_1{\tilde n}_2nk)\nonumber\\ 
&b(n_1n_2nk)^*=a({\tilde n}_1{\tilde n}_2nk)
F(n_1n_2nk)/{\bar \eta}({\tilde n}_1{\tilde n}_2nk)\nonumber\\ 
&b(n_1n_2nk)=a({\tilde n}_1{\tilde n}_2nk)^*
F(n_1n_2nk)/\eta({\tilde n}_1{\tilde n}_2nk)
\label{36}
\end{eqnarray}

Eqs. (\ref{35}) and (\ref{36}) define a possible symmetry when the 
set $(a,a^*)$ is replaced by the set $(b,b^*)$. Let us call it 
the AB symmetry. To understand whether it is indeed
a new symmetry, we should investigate when so defined $(b,b^*)$
operators satisfy the same commutation or anticommutation
relations as the $(a,a^*)$ operators, and whether the
operators $M^{ab}$ written in terms of $(b,b^*)$ have the
same form as in terms of $(a,a^*)$. 

As follows from Eqs. 
(\ref{30}) and (\ref{31}), the $b$-operators should 
satisfy either 
\begin{eqnarray}
&\{b(n_1n_2nk),b(n_1'n_2'n'k')^*\}=\nonumber\\
&Norm(n_1n_2nk)
\delta_{n_1n_1'}\delta_{n_2n_2'}\delta_{nn'}\delta_{kk'}
\label{37}
\end{eqnarray}
in the case of anticommutators or
\begin{eqnarray}
&[b(n_1n_2nk),b(n_1'n_2'n'k')^*]=\nonumber\\
&Norm(n_1n_2nk)
\delta_{n_1n_1'}\delta_{n_2n_2'}\delta_{nn'}\delta_{kk'}
\label{38}
\end{eqnarray}
in the case of commutators.

Now, as follows from Eqs. (\ref{24}),
(\ref{30}), (\ref{34}-\ref{36}), 
Eq. (\ref{37}) is satisfied if
\begin{equation}
\eta(n_1n_2nk) {\bar \eta}(n_1,n_2,nk)=(-1)^s
\label{39}
\end{equation}
At the same time, in the case of commutators it 
follows from Eqs. (\ref{24}), (\ref{31}) and 
(\ref{34}-\ref{36}) that Eq. (\ref{38}) is satisfied if
\begin{equation}
\eta(n_1n_2nk) {\bar \eta}(n_1,n_2,nk)=(-1)^{s+1}
\label{40}
\end{equation}

We now represent $\eta(n_1n_2nk)$ in the form 
\begin{equation}
\eta(n_1n_2nk)=\alpha f(n_1n_2nk)
\label{41}
\end{equation}
where $f(n_1n_2nk)$ should satisfy the condition
\begin{equation}
f(n_1n_2nk) {\bar f}(n_1,n_2,nk)=1
\label{42}
\end{equation}
Then $\alpha$ should be such that 
\begin{equation}
\alpha {\bar \alpha}=\pm (-1)^s
\label{43}
\end{equation}
where the plus sign refers to anticommutators and the minus
sign to commutators, respectively.
If the spin-statistics theorem is satisfied, i.e. we have 
anticommutators for
odd values of $s$ and commutators for even ones (this is the 
well known Pauli theorem in local quantum field theory 
\cite{Pauli}) then the r.h.s. 
of Eq. (\ref{43}) is equal to -1. 

Eq. (\ref{43}) is a consequence of the fact that our basis is
not normalized to one (see Ref. \cite{lev2} for discussion).
In the standard theory such a relation is 
impossible but if $\alpha\in F_{p^2}$, a solution of Eq. (\ref{43}) 
exists. Indeed, we can use the fact that any Galois field is 
cyclic with respect to multiplication \cite{VDW}. 
Let $r$ be a primitive root of $F_{p^2}$. This means that any
element of $F_{p^2}$ can be represented as a power of $r$. 
As mentioned in Sect. \ref{S2}, $F_{p^2}$ has only one
nontrivial automorphism which is defined as 
$\alpha\rightarrow {\bar \alpha}=\alpha^p$. 
Therefore if $\alpha =r^k$ then $\alpha{\bar \alpha}=
r^{(p+1)k}$. On the other hand, since $r^{(p^2-1)}=1$, we
conclude that $r^{(p^2-1)/2}=-1$. Therefore there exists at
least a solution with $k=(p-1)/2$. 

\section{Vacuum condition}
\label{Vacuum}

Although we have called the sets $(a,a^*)$ and 
$(b,b^*)$ annihilation and creation operators for
particles and antiparticles, respectively, it is not
clear yet whether these operators indeed can be treated
in such a way. 

In the standard approach, this can be
ensured by using the following procedure. One
requires the existence of the vacuum vector $\Phi_0$ such
that 
\begin{equation}
a(n_1n_2nk)\Phi_0=b(n_1n_2nk)\Phi_0=0\quad 
\forall\,\, n_1,n_2,n,k 
\label{v1}
\end{equation}
Then the elements 
\begin{equation}
\Phi_+(n_1n_2nk)=a(n_1n_2nk)^*\Phi_0\quad
\Phi_-(n_1n_2nk)=b(n_1n_2nk)^*\Phi_0
\label{v2}
\end{equation}
have the meaning of one-particle states for particles 
and antiparticles, respectively. 

However, if one requires the condition (\ref{v1})
in the GFQT then it is obvious from Eqs. (\ref{35})
and Eq. (\ref{36}), that the elements defined by 
Eq. (\ref{v2}) are null vectors. Note that in the
standard approach the AdS energy is always greater
than the mass while in the GFQT the AdS energy is not
positive definite. We can therefore try to modify
Eq. (\ref{v1}) as follows. Let us first break the 
set of elements $(n_1n_2nk)$ into two equal 
nonintersecting parts (defined later), 
$S_+$ and $S_-$, such that if $(n_1n_2nk)\in S_+$
then $({\tilde n}_1{\tilde n}_2nk)\in S_-$.
Then, instead of the condition (\ref{v1}) we 
require 
\begin{equation}
a(n_1n_2nk)\Phi_0=b(n_1n_2nk)\Phi_0=0\quad 
\forall\,\, (n_1,n_2,n,k)\in S_+ 
\label{v3}
\end{equation}
In that case the elements defined by Eq. (\ref{v2}) 
will indeed have the meaning of one-particle states
for $(n_1n_2nk)\in S_+$. 

By analogy with the standard approach, we can try
to define the set $S_+$ such that for the
corresponding values of $(n_1n_2nk)$ the AdS energy
$E=s+2(n+n_1+n_2+1)$ is positive. However, as already noted,
the meaning of positive and negative is not
quite clear in $F_p$. We can treat the AdS energy
as positive if all of the quantities $(nn_1n_2)$ are
much less than $p$ but in other cases such a treatment
would be problematic. We believe that in modern
physics there still exists a lack of understanding,
to what extent the positivity of energy is important.
For this reason our goal will be restricted to that
of constructing the set $S_+$ in a mathematically
consistent way. 
 
We will say that the AdS energy $E$ is positive if
$E$ is one of the values 1, 2,...$(p-1)/2$ and negative
if it is one of the values -1, -2,...$-(p-1)/2$. If $E$ is 
positive then we require that the corresponding
element $(n_1n_2nk)$ belongs to $S_+$, and if $E$ is
negative then the corresponding
element $(n_1n_2nk)$ belongs to $S_-$. The problem
arises with such elements that the corresponding value
of $E$ is equal to zero in $F_p$. Let us recall that
$E$ is the eigenvalue of $M^{05}=h_1+h_2$, the eigenvalue
of $h_1$ is equal to $E^{(1)}=1+s+n-k+2n_1$ and the
eigenvalue of $h_2$ is equal to $E^{(2)}=1+n+k+2n_2$.
The value of $E$ can be equal to zero in three cases:
$E^{(1)}$ is positive and $E^{(2)}$ is negative;
$E^{(1)}$ is negative and $E^{(2)}$ is positive;
$E^{(1)}=E^{(2)}=0$. We can require that in the
first case the corresponding element $(n_1n_2nk)$
belongs to $S_+$ and in the second case --- to $S_-$.
However, the third case is still problematic.

As follows from the results of Sects. \ref{S2} and
\ref{S3}, the case $E^{(1)}=0$ can occur only if
${\tilde n}_1=n_1$ where ${\tilde n}_1=N_1(n,k)-n_1$
and $N_1(n,k)$ is given by Eq. (\ref{21}).
Analogously the case $E^{(2)}=0$ can occur only if
${\tilde n}_2=n_2$ where ${\tilde n}_2=N_2(n,k)-n_2$.
Therefore the case $E^{(1)}=0$ can occur only if
$N_1(n,k)$ is even and $n_1=N_1(n,k)/2$.
Analogously, the case $E^{(2)}=0$ can occur only if
$N_2(n,k)$ is even and $n_2=N_2(n,k)/2$.

It is now clear that if the third case can occur
then the whole construction becomes inconsistent.
Indeed, since $b(n_1n_2nk)$ is proportional to
$a({\tilde n}_1 {\tilde n}_2 nk)^*$ then, if 
${\tilde n}_1=n_1$, ${\tilde n}_2=n_2$ and $\Phi_0$
is annihilated by both $a(n_1n_2nk)$ and 
$b(n_1n_2nk)$, it is also annihilated by both
$a(n_1n_2nk)$ and $a(n_1n_2nk)^*$. However this
contradicts Eqs. (\ref{30}) and (\ref{31}).   

Since $q_1=1+s$ and $q_2=1$ in the massless case, 
it follows from Eq. (\ref{21})
that if $s$ is even then $N_1(n,k)$ and $N_2(n,k)$
are either both even or both odd. Therefore in that
case we will necessarily have a situation when for
some values of $(nk)$, $N_1(n,k)$ and $N_2(n,k)$
are both even. In that case $E^{(1)}=E^{(2)}=0$ 
necessarily takes place for $n_1=N_1(n,k)/2$
and $n_1=N_1(n,k)/2$. Moreover, since for each 
$(nk)$ the number of all possible values of 
$(n_1n_2nk)$
is equal to $(N_1(n,k)+1)(N_2(n,k)+1$, this 
number is odd (therefore one cannot divide the set
of all possible values into the equal 
nonintersecting parts $S_+$ and $S_-$). 

On the other hand, if $s$ is odd then
for all the values of $(nk)$ we will necessarily have
a situation when either $N_1(n,k)$ is even and
$N_2(n,k)$ is odd or $N_1(n,k)$ is odd and
$N_2(n,k)$ is even. Therefore for each value of 
$(nk)$ the case $E^{(1)}=E^{(2)}=0$ is impossible, 
and the number of all possible values of $(n_1n_2nk)$  
is even.

We conclude that the condition (\ref{v3}) is
mathematically consistent only if $s$ is odd,
or in other words, if the particle spin in usual units
is half-integer. 

Although the interpretation of each operator from the set
$(a,a^*,b,b^*)$ as creation or annihilation one depends
on the way of breaking the elements $(n_1n_2nk)$ into
$S_+$ and $S_-$, the consistency requirement, that the
case $E^{(1)}=E^{(2)}=0$ should be excluded, does not
depend on the choice of $S_+$ and $S_-$. For this 
reason we believe, that the results of this section 
give a strong indication that in the massless case  
only particles with the half-integer spin can be 
elementary. Then as follows from the 
spin-statistics theorem \cite{Pauli}, 
massless elementary particles can be described only by 
anticommutation relations, i.e. they can be only 
fermions. However, since the spin-statistics theorem 
has not been proved in the GFQT yet, in the
subsequent sections we consider both anticommutators   
and commutators.   

\section{Compatibility of the AB symmetry with
representation operators}
\label{S5}

Let us consider the operators (\ref{32}) and
use the fact that in the modular case the trace of the
operators $M^{ab}$ is equal to zero (see Eq. (\ref{29})).   
Therefore, as follows from Eqs. (\ref{30}) and (\ref{31}), 
we can rewrite Eq. (\ref{32}) as
\begin{eqnarray}  
&M^{ab}=\mp \sum \{
M^{ab}(n_1'n_2'n'k',n_1n_2nk)\nonumber\\
&a(n_1n_2nk)a(n_1'n_2'n'k')^*/Norm(n_1n_2nk)\}
\label{44}
\end{eqnarray}
where the minus sign refers to anticommutators and the plus sign -
to commutators. Using Eqs. (\ref{34}-\ref{36}) and 
(\ref{41}-\ref{43}), we then obtain
\begin{eqnarray}  
&M^{ab}=-\sum\{
M^{ab}(n_1'n_2'n'k',n_1n_2nk)
f(n_1'n_2'n'k'){\bar f}(n_1n_2nk)\nonumber\\
&b({\tilde n}_1{\tilde n}_2nk)^*
b({\tilde n}_1'{\tilde n}_2'n'k')/
[F({\tilde n}_1'{\tilde n}_2'n'k')G(k)]\}=\nonumber\\
&-\sum\{ M^{ab}({\tilde n}_1{\tilde n}_2nk,
{\tilde n}_1'{\tilde n}_2'n'k')
f({\tilde n}_1{\tilde n}_2nk)
{\bar f}({\tilde n}_1'{\tilde n}_2'n'k')\nonumber\\
&b(n_1'n_2'n'k')^*b(n_1n_2nk)/[F(n_1n_2nk)G(k')]\}
\label{45}
\end{eqnarray}
in both cases.

We first consider the AdS energy operator which is diagonal.
As follows from Eq. (\ref{22}), in the massless case the 
matrix elements of the $M^{05}$ operator are given by
\begin{equation}
M^{05}(n_1'n_2'n'k'n_1n_2nk)=
(2+s+2n+2n_1+2n_2)\delta_{n_1n_1'}
\delta_{n_2n_2'}\delta_{nn'}\delta_{kk'}
\label{46}
\end{equation}
Therefore the operator (\ref{32}) in this case can be written
as
\begin{eqnarray}  
&M^{05}=\sum_{n_1n_2nk} \{
[s + 2(n+n_1+n_2+1)]a(n_1n_2nk)^*\times\nonumber\\
&a(n_1n_2nk)/Norm(n_1n_2nk)\}
\label{47}
\end{eqnarray}
At the same time, as follows from Eqs. (\ref{42}), (\ref{45}), 
(\ref{46}) and the definition of the transformation 
$n_1\rightarrow {\tilde n}_1,\,\,n_2\rightarrow {\tilde n}_2$ 
(see Sect. \ref{S4})
\begin{eqnarray}  
&M^{05}=\sum_{n_1n_2nk}\{
[s + 2(n+n_1+n_2+1)]b(n_1n_2nk)^*\times\nonumber\\
&b(n_1n_2nk)/Norm(n_1n_2nk)\}
\label{48}
\end{eqnarray}

In Eqs. (\ref{47}) and (\ref{48}), the sum is taken over
all the values of $(n_1n_2nk)$ relevant to the particle modular IR.
At the same time, for the correspondence with the standard
case, we should consider only the values of the $(n_1n_2n)$ which
are much less than $p$ (see Refs. \cite{lev1,lev2}). The 
derivation of Eq. (\ref{48}) demonstrates that the contribution of
those $(n_1n_2n)$ originates from such a
contribution of $(n_1,n_2)$ to Eq. (\ref{47}) that
$({\tilde n}_1,{\tilde n}_2)$ are much less than $p$.
In this case the $(n_1,n_2)$ are comparable to $p$. Therefore,
if we consider only such states that the $(n_1n_2n)$ in the
$a$ and $b$ operators are much less than $p$ then the AdS 
Hamiltonian can be written in the form
\begin{eqnarray}  
&M^{05}=\sum_{n_1n_2nk}' \{
[s + 2(n+n_1+n_2+1)][a(n_1n_2nk)^*\times\nonumber\\
&a(n_1n_2nk)+b(n_1n_2nk)^*b(n_1n_2nk)]/Norm(n_1n_2nk)\}
\label{49}
\end{eqnarray}
where $\sum_{n_1n_2nk}'$ means that the sum is taken only
over the values of the $(n_1n_2nk)$ which are much less than $p$.
In this expression the contributions of particles and
antiparticles are written down explicitly and the corresponding
standard AdS Hamiltonian is positive definite.

The above results show that as far as the operator $M^{05}$ is
concerned, Eq. (\ref{35}) indeed defines a new symmetry since
$M^{05}$ has the same form in terms of $(a,a^*)$ and $(b,b^*)$
(compare Eqs. (\ref{47}) and (\ref{48})). Note that we did
not assume that the theory was C-invariant (in the standard
theory C-invariance can be defined as the transformation
$$a(n_1n_2nk)\leftrightarrow b(n_1n_2nk)).$$
It is well known that C-invariance is not a fundamental symmetry.
In the standard theory only CPT-invariance is
fundamental since, according to the famous CPT-theorem \cite{GLR}, 
any local Poincare invariant theory is automatically CPT-invariant. 
Our assumption is that Eq. (\ref{35}) defines a fundamental
symmetry in the GFQT. To understand its properties one has to 
investigate not only $M^{05}$ but other representation generators 
as well.  

By analogy with the case of the operator $M^{05}$, it is easy
to show that at the same conditions, the operators $h_1$ and
$h_2$ have the same form in terms of $(a,a^*)$ and $(b,b^*)$.

Consider now the operator $a_1"$ (see Sect. \ref{S2}).
As follows from its definition, its matrix elements are
given by
\begin{equation}
a_1"(n_1'n_2'n'k'n_1n_2nk)=\delta_{n_1,n_1'-1}
\delta_{n_2n_2'}\delta_{nn'}\delta_{kk'}
\label{50}
\end{equation}
and therefore, as follows from Eq. (\ref{32}),
the secondly quantized form of $a_1"$ is
\begin{eqnarray}
a_1"=\sum_{n_1=0}^{N_1-1}\sum_{n_2nk} \{ a(n_1+1,n_2nk)^*
a(n_1n_2nk)/Norm(n_1n_2nk)\}
\label{51}
\end{eqnarray}
We have to prove that in terms of $(b,b^*)$ this operator
has the same form, i.e.
\begin{eqnarray}
a_1"=\sum_{n_1=0}^{N_1-1}\sum_{n_2nk} \{ b(n_1+1,n_2nk)^*
b(n_1n_2nk)/Norm(n_1n_2nk)\}
\label{52}
\end{eqnarray}
As follows from Eqs. (\ref{45}) and (\ref{50}), Eq. (\ref{52})
is indeed valid if 
\begin{equation}
f(n_1n_2nk){\bar f}(n_1-1,n_2nk)=-1
\label{53}
\end{equation} 

Since the action of the operator $a_1'$ can be written as
$$a_1'e(n_1n_2nk)=a_1'a_1"e(n_1-1,n_2nk)$$
then, as follows from Eq. (\ref{10}), the matrix
elements of the operator $a_1'$ are given by
\begin{equation}
a_1'(n_1'n_2'n'k'n_1n_2nk)=n_1(Q_1(n,k)+n_1-1)\delta_{n_1,n_1'+1}
\delta_{n_2n_2'}\delta_{nn'}\delta_{kk'}
\label{54}
\end{equation}
Therefore, as follows from Eq. (\ref{32}),
the secondly quantized form of this operator is
\begin{eqnarray}
&a_1'=\sum_{n_1=1}^{N_1}\sum_{n_2nk} \{ n_1(Q_1(n,k)+n_1-1)
a(n_1-1,n_2nk)^*\nonumber\\
&a(n_1n_2nk)/Norm(n_1n_2nk)\}
\label{55}
\end{eqnarray}
By analogy with the proof of Eq. (\ref{52}), one
can prove that in terms of $(b,b^*)$ this operator
has the same form, i.e.
\begin{eqnarray}
&a_1'=\sum_{n_1=1}^{N_1}\sum_{n_2nk} \{n_1(Q_1(n,k)+n_1-1)
b(n_1-1,n_2nk)^*\nonumber\\
&b(n_1n_2nk)/Norm(n_1n_2nk)\}
\label{56}
\end{eqnarray}
if
\begin{equation}
f(n_1n_2nk){\bar f}(n_1+1,n_2nk)=-1
\label{57}
\end{equation} 

Analogously we can prove that the secondly quantized operators 
$a_2"$ and $a_2'$ also have the same form in terms of $(a,a^*)$
and $(b,b^*)$ if
\begin{eqnarray}
&f(n_1n_2nk){\bar f}(n_1,n_2+1,nk)=-1\nonumber\\ 
&f(n_1n_2nk){\bar f}(n_1,n_2-1,nk)=-1
\label{59}
\end{eqnarray} 

As follows from Eqs. (\ref{42}), (\ref{53}, (\ref{57}) and
(\ref{59}), the function $f(n_1n_2nk)$ necessarily has the
form
\begin{equation}
f(n_1n_2nk)=(-1)^{n_1+n_2}f(n,k)
\label{60}
\end{equation}
where the function $f(n,k)$ should satisfy the condition
\begin{equation}
f(n,k){\bar f}(n,k)=1
\label{61}
\end{equation}

The next step is to investigate whether the remaining 
operators $(b',b",L_+,L_-)$ have the same 
form in terms of $(a,a^*)$ and $(b,b^*)$. We discuss
the operator $b'$ since computations with the
other operators are analogous (and simpler).

As follows from Eqs. (\ref{4}) and (\ref{20}),
\begin{eqnarray}
&b'e(n_1n_2nk)=b'(a_1")^{n_1}(a_2")^{n_2}e_{nk}=
(a_1")^{n_1}b'(a_2")^{n_2}e_{nk}+\nonumber\\
&n_1(a_1")^{n_1-1}L_+(a_2")^{n_2}e_{nk}=
(a_1")^{n_1}(a_2")^{n_2}b'e_{nk}+\nonumber\\
&n_2(a_1")^{n_1}(a_2")^{n_2-1}L_-e_{nk}+
n_1(a_1")^{n_1-1}(a_2")^{n_2}L_+e_{nk}+\nonumber\\
&n_1n_2(a_1")^{n_1-1}(a_2")^{n_2-1}b"e_{nk}
\label{62}
\end{eqnarray}
By using Eq. (\ref{12}) we can express the action of
the $(b',b",L_+,L_-)$ operators on the minimal vectors
in terms of the $B$ operators:
\begin{eqnarray}
&b"=B^{++}+a_1"B^{-+}(h_1-1)^{-1}+a_2"B^{+-}(h_2-1)^{-1}+\nonumber\\
&a_1"a_2"B^{--}[(h_1-1)(h_2-1)]^{-1}\quad L_+=
B^{+-}+a_1"B^{--}(h_1-1)^{-1}\nonumber\\
&L_-=B^{-+}+a_2"b'(h_2-1)^{-1}\quad b'=B^{--}
\label{63}
\end{eqnarray}
and then use Eq. (\ref{23}). 

In such a way we can 
explicitly compute $b'e(n_1n_2nk)$ in Eq. (\ref{62}),
find the matrix elements of $b'$ by using Eq. (\ref{28})
and write the operator $b'$ in the secondly quantized form
by using Eq. (\ref{32}). The result is
\begin{eqnarray}
&b'=\sum_{n=0}^{n_{max}-1}\sum_{k=0,s}\sum_{n_1=1}^{N_1}
\sum_{n_2=1}^{N_2}\{(n_1n_2)\nonumber\\
&a(n_1-1,n_2-1,n+1,k)^*a(n_1n_2nk)/Norm(n_1n_2nk)\} +\nonumber\\
&\sum_{n=1}^{n_{max}}\sum_{k=0,s}\sum_{n_1=0}^{N_1}
\sum_{n_2=0}^{N_2}\{ (n+s-k+n_1)(n+k+n_2)\nonumber\\
&a(n_1n_2,n-1,k)^*a(n_1n_2nk)/Norm(n_1n_2nk)\} +\nonumber\\
&\sum_{k=0}^{s-1}\sum_{n_1=0}^{N_1}
\sum_{n_2=1}^{N_2}\{ [n_2(s-k+n_1)/(s-k)]\nonumber\\
&a(n_1,n_2-1,0,k+1)^*a(n_1n_20k)/Norm(n_1n_20k)\}+\nonumber\\
&\sum_{k=1}^{s}\sum_{n_1=1}^{N_1}
\sum_{n_2=0}^{N_2}\{ n_1(n_2+k)(s+1-k)\nonumber\\
&a(n_1-1,n_2,0,k-1)^*a(n_1n_20k)/Norm(n_1n_20k)\}
\label{64}
\end{eqnarray}
where (see Sect. \ref{S3}) $n_{max}=p-1-s$, $N_1=N_1(n,k)$ 
and $N_2=N_2(n,k)$.

The next step is to express the $(a,a^*)$ operators in terms
of the $(b,b^*)$ operators by using Eqs. (\ref{35}) and 
(\ref{36}), and use Eqs. (\ref{34}), (\ref{41}), (\ref{43})
and (\ref{60}). The result is as follows. 
Eq. (\ref{64}) has the same form in terms of $(a,a^*)$
and $(b,b^*)$ only if 
\begin{equation}
f(n,k)=c (-1)^n 
\label{69}
\end{equation}
where $c$ is any constant such that $c\bar{c}=1$.

Analogous computations for the operators $(b"L_+L_-)$ 
show that if Eq. (\ref{69}) is satisfied then they have
the same form in terms of $(a,a^*)$ and $(b,b^*)$.
Therefore, as follows from Eqs. (\ref{41}) and (\ref{60}), the
final solution for $\eta(n_1n_2nk)$ is
\begin{equation}
\eta(n_1n_2nk)=\alpha f(n_1n_2nk)\quad f(n_1n_2nk)=(-1)^{n_1+n_2+n}
\label{70}
\end{equation}
where $\alpha$ satisfies Eq. (\ref{43}).

We have proved that the AB symmetry defined by Eq. 
(\ref{35}) is indeed a fundamental symmetry in the GFQT
(at least for massless elementary particles).  

\section{Problem of existence of neutral elementary particles}
\label{S6}

Suppose now that the particle in question is neutral, i.e.
the particle coincides with its antiparticle. On the language
of the operators $(a,a^*)$ and $(b,b^*)$ this means that these
sets are the same, i.e. $a(n_1n_2nk)=b(n_1n_2nk)$ and
$a(n_1n_2nk)^*=b(n_1n_2nk)^*$. As a consequence, Eq. 
(\ref{35}) has now the form  
\begin{equation}
a(n_1n_2nk)^*=\eta(n_1n_2nk) a({\tilde n}_1{\tilde n}_2nk)/
F({\tilde n}_1{\tilde n}_2nk) 
\label{71}
\end{equation}
and therefore 
\begin{equation}
a(n_1n_2nk)={\bar \eta}(n_1n_2nk) a({\tilde n}_1{\tilde n}_2nk)^*/
F({\tilde n}_1{\tilde n}_2nk) 
\label{72}
\end{equation}

As follows from Eqs. (\ref{41}) and (\ref{43}),
these expressions are
compatible with each other only if
\begin{equation}
f(n_1n_2nk){\bar f}({\tilde n}_1,{\tilde n}_2,nk)=\pm 1
\label{73}
\end{equation}
where the plus sign refers to anticommutators and the minus
sign to commutators, respectively. Therefore the problem arises
whether Eqs. (\ref{70}) and (\ref{73}) are compatible with
each other. As follows from Eq. (\ref{21}), (\ref{60}),
(\ref{61}) and the definition of the transformations 
$n_j\rightarrow {\tilde n}_j$ (see Sect. \ref{S4}) 
\begin{equation}
f(n_1n_2nk){\bar f}({\tilde n}_1{\tilde n}_2nk)=(-1)^s
\label{74}
\end{equation}
By comparing Eqs. (\ref{73}) and (\ref{74}) we conclude
that they are incompatible with each other if the  
spin-statistics theorem is satisfied. 
Therefore massless particles in the GFQT cannot be
elementary but only composite. 

\section{Discussion}
\label{S7}

In the present paper we have considered massless IRs in
a quantum theory based on a Galois field (GFQT). One of
the crucial differences between the GFQT and the 
standard theory is that in the GFQT a particle and its 
antiparticle represent different states of the same 
object. As a consequence, the annihilation and creation
operators for a particle and its antiparticle can be 
directly expressed in terms of each other. This imposes 
additional restrictions on the structure of the theory. 
In particular, Eq. (\ref{35}) defines a new symmetry which
has no analog in the standard theory. We have shown in
Sect. \ref{S5} that this is indeed a symmetry in the
massless case since the representation operators have 
the same form in terms of annihilation and creation 
operators for particles and antiparticles. It has been
also shown in Sect. \ref{Vacuum} that the new symmetry
is compatible with the vacuum condition only for
particles with the half-integer spin (in conventional
units). As a consequence,
as shown in Sect. \ref{S6}, the existence of massless 
neutral elementary particles in the GFQT is incompatible
with the spin-statistics theorem.    
It will be shown in a separate paper that these results
can be extended to the massive case as well.

Is it natural 
that the requirement about the normal connection 
between spin and
statistics excludes the existence of neutral elementary
particles? If there is no restriction
imposed by the spin-statistics theorem then we cannot
exclude the existence of neutral elementary particles in the
GFQT. However, such an existence seems to be rather unnatural.
Indeed, since one modular IR simultaneously describes a 
particle and its antiparticle, the AdS energy operator 
necessarily contains the contribution of
the both parts of the spectrum, corresponding to the particle
and its antiparticle (see Eq. (\ref{49})). If a particle
were the same as its antiparticle then Eq. (\ref{49}) would 
contain two equal contributions and thus the value of the AdS
energy would be twice as big as necessary. 

Although the conclusion about the nonexistence
of neutral elementary particles has been
made for both bosons and fermions, it is obvious that
the case of bosons is of greater importance. 
A possibility that the photon is composite has been 
already discussed in the literature. For 
example, in Ref. \cite{FF} a model 
where the photon is composed of two Dirac singletons 
\cite{DiracS} has been investigated. However, in the 
framework of the standard theory, the compositeness of 
the photon is only a possible (and attractive) scenario 
while in the GFQT this is inevitable.

It is well
known that the standard local quantum field theory (LQFT)
has achieved very impressive success in comparing theory 
and experiment. In particular, quantum 
electrodynamics and the electroweak theory are based
on the assumption that the photon is the elementary
particle. For this reason one might doubt whether our
conclusion has any relevance to physics. At the same time
the LQFT has several well known 
drawbacks and inconsistencies. The majority of physicists
believes that \cite{Wein} the LQFT
should be $'$taken as is$'$, but at the same
time it is a $'$low energy approximation to a
deeper theory that may not even be a field theory, but 
something different like a string theory$'$ \cite{Wein}.  

We argued in \cite{lev1,lev2} that the future quantum
physics will be based on a Galois field. In that case the
theory does not contain actual infinity, all operators are
well defined, divergencies cannot exist in principle etc.
We believe however that not only this makes the GFQT very 
attractive.

For centuries, scientists and philosophers have been trying to
understand why mathematics is so successful in explaining
physical phenomena (see e.g. Ref. \cite{Wigner1}). However,
such a branch of mathematics as number theory and, in 
particular, Galois fields, have practically no implications 
in physics.
Historically, every new physical theory usually involved
more complicated mathematics. The standard mathematical
tools in modern quantum theory are differential and
integral equations, distributions, analytical functions,
representations of Lie algebras in Hilbert spaces etc.
At the same time, very impressive results of number theory
about properties of natural numbers (e.g. the Wilson theorem)
and even the notion of primes are not used at all!
The reader can easily notice that the GFQT involves only 
arithmetic of Galois fields (which are even simpler than
the set of natural numbers). The very possibility that 
the future quantum theory could be formulated in such a way,
is of indubitable interest.


\begin{thebibliography}{99}
\bibitem{Wigner} E.P. Wigner, Ann. Math. {\bf 40}, 149 (1939).
\bibitem{DirNobel} P.A.M. Dirac, in {\it The World Treasury
of Physics, Astronomy and Mathematics}, p. 80, Timothy Ferris ed.,
(Little Brown and Company, Boston-New York-London, 1991).
\bibitem{lev1} F.M. Lev, Yad. Fiz. {\bf 48}, 903 (1988); 
J. Math. Phys. {\bf 30}, 1985 (1989); J. Math. Phys. {\bf 34},
490 (1993).
\bibitem{lev2} F.M. Lev, hep-th/0206078.
\bibitem{VDW} B.L. Van der Waerden, {\it Algebra I}, 
(Springer-Verlag, Berlin Heidelberg New York, 1967);
K. Ireland and  M. Rosen, {\it A Classical 
Introduction to Modern Number Theory, Graduate Texts in 
Mathematics-87}, (New  York  - Heidelberg - Berlin: Springer, 1987); 
H. Davenport, {\it The Higher Arithmetic}, 
(Cambridge University Press, 1999).
\bibitem{FrPa} E.M. Friedlander and B.J. Parshall, Bull. Am.
Math. Soc. {\bf 17}, 129 (1987).
\bibitem{Fronsdal} C. Fronsdal, Rev. Mod. Phys. {\bf 37}, 221 
(1965). 
\bibitem{Evans} N.T. Evans, J. Math. Phys. {\bf 8}, 170 (1967). 
\bibitem{Braden} B. Braden, Bull. Amer. Math. Soc. 
{\bf 73}, 482 (1967);
{\it Thesis, Univ. of Oregon}, (Eugene, OR, 1966).
\bibitem{Zass} H. Zassenhaus, Proc. Glasgow Math. Assoc. 
{\bf 2}, 1 (1954).
\bibitem{IW} E. Inonu and E.P. Wigner, Nuovo Cimento, {\bf IX}, 
705 (1952).
\bibitem{AB} Yu.V. Novozhilov, 
{\it An Introduction to Elementary Particle
Theory} (Nauka, Moscow, 1972).
\bibitem{Wein} S. Weinberg, {\it Quantum Theory of Fields}, 
(Cambridge, Cambridge University Press, 1995).
\bibitem{Pauli} W. Pauli, Phys. Rev. {\bf 58}, 116 (1940).
\bibitem{GLR} W. Pauli, in {\it N. Bohr and the Development of
Physics}, (Pergamon Press, London, 1955); G. Gravert, G. Luders and
G. Rollnik, Fortschr. Phys. {\bf 7}, 291 (1959).
\bibitem{FF} M. Flato and C. Fronsdal, Lett. Math. Phys.
{\bf 2}, 421 (1978); L. Castell and W. Heidenreich, Phys. Rev. 
{\bf D24}, 371 (1981); C. Fronsdal, Phys. Rev. 
{\bf D26}, 1988 (1982).
\bibitem{DiracS} P.A.M. Dirac, J. Math. Phys. {\bf 4}, 901 (1963).
\bibitem{Wigner1} E.P. Wigner, in {\it The World Treasury
of Physics, Astronomy and Mathematics}, p. 526, Timothy Ferris ed.,
(Little Brown and Company, Boston-New York-London, 1991).
\end{thebibliography}
\end{document}